\documentclass[aps,preprint,showkeys,showpacs]{revtex4}

\usepackage{amsmath}
\usepackage{amsfonts}
\usepackage{amssymb}
\usepackage{graphicx}

\begin{document}

\title{High energy astrophysical neutrino flux \\and modified dispersion relations}

\author{J.L. \surname{Bazo}}
\email{jlbazo@pucp.edu.pe}
\author{M. \surname{Bustamante}}
\email{mbustamante@pucp.edu.pe}
\author{A.M. \surname{Gago}}
\email{agago@pucp.edu.pe}
\affiliation{Pontificia Universidad Cat\'{o}lica del Per\'{u}, Depto.~de Ciencias, 
Secci\'on F\'{\i}sica \\ Apartado postal 1761, Lima, Peru }

\author{O.G. \surname{Miranda}}
\email{omr@fis.cinvestav.mx}
\affiliation{Centro de Investigaci\'on y de Estudios Avanzados del IPN,
Depto.~de F\'isica \\ Apartado postal 14-740, M\'exico 07000 D.F. Mexico}

\begin{abstract}
Motivated by the interest in searches for violation of CPT invariance,
we study its possible effects in the flavour ratios of high-energy
neutrinos coming from cosmic accelerators. In particular, we focus on the 
effect of an energy-independent new physics contribution
to the neutrino flavour oscillation phase and explore whether it is
observable in future detectors. 
Such a contribution could be related not only to CPT
violation but also to a nonuniversal coupling of neutrinos to a
torsion field. We conclude that this extra phase contribution only
becomes observable, in the best case, at energies greater than
$10^{16.5}$ GeV, which is about five orders of magnitude higher 
than the most energetic cosmological neutrinos to be detected 
in the near future. 
Therefore, if 
these effects are present only in the oscillation phase, they are
going to be unobservable, unless a new mechanism or source capable to
produce neutrinos of such energy were detected.
\end{abstract}


\maketitle

\section{Introduction}\label{Section_Introduction}

Experimental evidence has confirmed that flavour transitions are the
solution to the former so-called solar and atmospheric neutrino
deficit problems\cite{SNO_05,Eguchi:2002dm,Maltoni:2004ei}.  
Further evidence was provided by experiments performed with 
neutrinos generated in particle accelerators and nuclear reactors, such as
KamLAND\cite{Collaboration:2008ee} and K2K\cite{Ahn:2006zza}. The
mechanism responsible for these transitions requires neutrinos to be
massive: the probability of a flavour transition is oscillatory, with
oscillation length $\lambda^{std}\equiv4\pi E/\Delta m^2$, where $E$
is the neutrino energy and $\Delta m^2$ is the difference of the
squared masses of the different neutrino mass eigenstates.  However,
even though this mass-driven mechanism is the dominant one in the
energy regimes that have been explored experimentally (MeV-TeV), there
is still the possibility that alternative mechanisms contribute to the
flavour transitions in a subdominant manner, which perhaps can
manifest at higher energies.

Although these alternative mechanisms involving new physics (NP) are
able to produce flavour transitions, it is known that none of them can
explain the combined data from atmospheric, solar, accelerator, and
reactor neutrino experiments performed in the MeV-TeV range, unlike
the pure $\Delta m^2$ oscillation
mechanism\cite{GonzalezGarcia_04,Fogli_99,Battistoni_05,Fogli_07}.
Some of these alternatives\cite{Hooper_05} are the violation of the
equivalence principle (VEP), of Lorentz
invariance\cite{Coleman_99,Mattingly_05} (VLI), of CPT invariance
(VCPT), the non-universal coupling of neutrinos to a space-time
torsion field (NUCQ), decoherence during the neutrino's trip, and
non-standard interactions (NSI).

Typically, these mechanisms result in oscillation lengths
$\lambda^{NP}$ that have a different dependence on $E$, usually
expressed as a power-law, $\lambda^{NP}\sim E^n$, with the value of
$n$ depending on which mechanism is being considered: for instance,
$n=0$ for VCPT and NUCQ and $n=1$ for VEP and VLI, while with $n=-1$
the standard $\Delta m^2$ oscillations are recovered. Atmospheric
events from Super-Kamiokande (SK)\cite{Fogli_99} were used to find the
value $n=-0.9\pm0.4$ at $90\%$ C.L., thus confirming the dominance of
the $\Delta m^2$ oscillation mechanism and forcing any other
mechanisms to be subdominant, at least within the energy range and
pathlength considered in said analysis.

So far, searches for NP effects in neutrino oscillations have been
limited to energies ranging from a few MeV to a few
GeV\cite{GonzalezGarcia_04,Fogli_99,Battistoni_05,Fogli_07} and have
turned out negative.  However, proposals for analyses of atmospheric
neutrinos with energies of up to $10^4$ TeV in second-generation
neutrino detectors such as IceCube\cite{GonzalezGarcia_05} and
ANTARES\cite{Morgan_07} are being considered. Due to the energy
dependence of the oscillation lengths, the oscillation phases scale as
$\left(2\pi L/\lambda^{NP}\right)/\left(2\pi/\lambda^{std}\right)\sim
E^{1-n}$, that is, the relative dominance of the NP contribution grows
with the neutrino energy provided that $n\le0$, so that the
observation of very energetic neutrinos \--such as the ones expected
from presumed cosmic accelerators like active galaxies and gamma ray
bursts\-- would offer a means to establish whether the $\Delta m^2$
oscillation mechanism is still the dominant one at high energies or to
otherwise set stronger bounds on the NP parameters.

In this work, we have introduced the aforementioned new physics
through the use of a modified dispersion relation, and focused our
analysis on the case of $n=0$, corresponding, as we will see, to an
energy-independent NP contribution to the neutrino flavour oscillation
phase. We have calculated the proportion of each flavour arriving at
Earth from distant cosmic accelerators and explored how it is affected
by the parameters that control the new physics, and whether these
effects are observable at all.

The outline of the paper is as follows. Section
\ref{Section_ModificationsQG} describes how the NP arising from a
modified dispersion relation affects the flavour-transition
probability for neutrinos that travel cosmological distances.  In
Section \ref{Section_Observability} we explore the case of an
energy-independent NP contribution and its effects on the flavour
ratios. Finally, in Section \ref{Section_Conclusions}, we present our
conclusions.

\section{Flavour-transition probability in the presence of a modified 
 dispersion relation}\label{Section_ModificationsQG}

The NP effects can modify flavour transitions in two ways\cite{Morgan_07}: 
by transforming both the oscillation length and the
neutrino mixing angles or by altering only the first. The former case
occurs, for instance, when considering the low energy phenomenological
model of string theory, known as 'Standard Model Extension'\cite{Hooper_05} 
and has been examined using SK and K2K data\cite{GonzalezGarcia_04,GonzalezGarcia_05}. 
The second case can be achieved by considering a modified dispersion
relation which departs from the well-known formula $E^2 = p^2+m^2$.
Because we wish to explore whether solely effects on the phase are
observable at high energies, we follow this second alternative and
consider the following modified dispersion
relation\cite{AmelinoCamelia_05}, which allows us to study the
contributions of NP effects in a model-independent way:
\begin{equation}\label{ModDispRel}
E^2 = p^2 + m^2 + \eta' p^2\left(\frac{E}{m_P}\right)^\alpha = p^2 + m^2 + \eta p^2 E^\alpha ~,
\end{equation}
where $m_{P} \simeq 10^{19}$ GeV is the Planck mass, $\eta'= \eta
m_P^\alpha$ is an adimensional parameter that controls the strength of
the NP effects and,
following the literature, $\alpha$ has been chosen to be of integer
value.
Such a dispersion relation assumes that the scale of NP effects is the
Planck scale where, according to theories of quantum gravity,
space-time might become ``foamy''. 
Eq. (\ref{ModDispRel}) was recently\cite{Morgan_07} used to
predict the sensitivity of the ANTARES neutrino telescope to NP
effects in the high-energy atmospheric neutrino flux.

We now derive the flavour-transition probability in the presence of NP
effects, for neutrinos that propagate over a cosmological
distance. Flavour transitions arise as a consequence of the
fact that flavour eigenstates $\vert\nu_\alpha\rangle$
($\alpha=e,\mu,\tau$) are not also mass eigenstates
$\vert\nu_i\rangle$ ($i=1,2,3$), but rather a linear combination
of them, i.e. $\vert\nu_\alpha\rangle=\sum_{i=1}^3 U_{\alpha i}^\ast
\vert\nu_i\rangle$, with $U_{\alpha i}^\ast$ elements of the 
neutrino mixing matrix. 

Using the standard dispersion relation, it is a common
procedure to derive an approximate expression for the momentum of the
$i$-th neutrino mass eigenstate,
\begin{equation}\label{MomentumStd}
p_{i} = \sqrt{E^2 - m_i^2} \simeq E - \frac{m_i^2}{2E} ~,
\end{equation}
where $m_i$ is the mass of the neutrino and $E$ is its energy, and are such that, at the
energies that we have considered, $m_i\ll E$. From this equation 
we obtain the usual expression for the momenta difference:
\begin{equation}\label{MomentumDiffStd}
\Delta p_{ij} \equiv p_j - p_i = \frac{\Delta m_{ij}^2}{2E} ~.
\end{equation}
In accordance with the latest bounds obtained from global
fits\cite{Schwetz:2008er}, we have set the three mixing angles that
parametrise $U$ to $\sin^2\left(\theta_{12}\right) = \sqrt{0.304}$,
$\theta_{13} = 0$ and $\theta_{23} = \pi/4$. The mass-squared
differences have been set to $\Delta m_{21}^{2} = 8.0 \times
10^{-5}~\text{eV}^{2}$ and $\Delta m_{32}^{2} = 2.5 \times
10^{-3}~\text{eV}^{2}$, and we have assumed a normal mass hierarchy
(i.e. $m_3 > m_1$), so that $\Delta m_{31} = \Delta m_{32} + \Delta
m_{21}$. The probability that a neutrino created with flavour $\alpha$
is detected as having flavour $\beta$ after having propagated a
distance $L$ in vacuum is given by\cite{Kayser:2005cd}
\begin{equation}\label{ProbKayserStdB}
P_{\nu_\alpha\rightarrow\nu_\beta}\left(E, L\right) = \delta_{\alpha\beta} 
- 4 \sum_{i>j} \text{Re}\left(J_{ij}^{\alpha\beta}\right) \sin^2\left(\frac{\Delta p_{ij}}{2}L\right) ~,
\end{equation}
where $J_{ij}^{\alpha\beta} \equiv U_{\alpha i}^\ast U_{\beta i} U_{\alpha j} U_{\beta j}^\ast$. Since $\theta_{13} = 0$, $U$ is a real matrix, independent of the CP-violation phase, $\delta_{CP}$.

In the case of the modified dispersion relation of
Eq.~(\ref{ModDispRel}) we can also find an expression for the momenta
difference. To first order in $\eta_i$, and discarding terms higher
than second power in $m_i$ or involving $\eta_i m_i^2$, we obtain
\begin{equation}
p_{i} \simeq E - \frac{m_{i}^2}{2E} - \frac{\eta_i E^n}{2} ~,
\end{equation}
with $n\equiv\alpha+1$, and hence
\begin{equation}\label{MomentumDiffNP}
\Delta \tilde{p}_{ij} = \frac{\Delta m_{ij}^2}{2E} + \frac{\Delta \eta_{ij}^{\left(n\right)} E^n}{2} ~, 
\end{equation}
where $\Delta
\eta_{ij}^{\left(n\right)}\equiv\eta_i^{\left(n\right)}-\eta_j^{\left(n\right)}$,
for the NP mechanism with an $E^n$ energy dependence. 
Note that it is necessary that the $\eta_i$ have different values for
different mass eigenstates in order to have a nonzero NP contribution
to the momenta difference. The corresponding oscillation probability
is Eq. (\ref{ProbKayserStdB}) with $\Delta p_{ij} \rightarrow \Delta
\tilde{p}_{ij}$; hence, the NP affects solely the oscillation phase,
but not its amplitude.

Since $L \gg 1$ for high-energy astrophysical neutrinos,
$\sin^2\left(\Delta p_{ij} L /2\right)$ is a rapidly oscillating
function and so, due to the limited energy resolution of neutrino
telescopes, the average flavour-transition probability is sometimes
used instead, i.e.
\begin{equation}\label{ProbAvg}
 \langle P_{\nu_\alpha\rightarrow\nu_\beta} \rangle = \sum_i \lvert U_{\alpha i} \rvert^2 \lvert U_{\beta i} \rvert^2 ~.
\end{equation}

Let $\langle P_{\nu_\alpha\rightarrow\nu_\beta} \rangle^{std}$ be the
standard average probability, that is, when there are no NP effects
present. If the extra term in $\Delta \tilde{p}_{ij}$ has $n > 0$, its
effect will be that, at high energies,
$P_{\nu_\alpha\rightarrow\nu_\beta}$ will oscillate even more rapidly
with energy, but still around the same mean value $\langle
P_{\nu_\alpha\rightarrow\nu_\beta} \rangle^{std}$. If $n<0$, the
oscillations will continue up to high energies, also around the same
mean, and at a high enough value, when $\Delta p_{ij} = \Delta
m_{ij}^2/2E \rightarrow 0$, the probability will tend to
zero. However, if $n=0$, then the extra term in $\Delta
\tilde{p}_{ij}$ is energy-independent and so when $\Delta p_{ij}
\rightarrow 0$, $P_{\nu_\alpha\rightarrow\nu_\beta}$ becomes constant,
but different from zero due to the existence of the extra
term. Furthermore, when this happens, and depending on the values of
the $\Delta \eta_{ij}^{\left(0\right)}$, it is in principle possible
for the constant probability to be different from $\langle
P_{\nu_\alpha\rightarrow\nu_\beta} \rangle^{std}$. Such a nonzero,
constant probability at high energies could therefore be interpreted
as being due to the contribution from energy-independent new
physics. We will focus on this possibility.

Although our analysis of NP effects using Eq.~(\ref{ModDispRel}) is
model-independent, $\Delta \eta^{\left(0\right)}$ takes a different
form depending on the particular mechanism being
considered\cite{GonzalezGarcia_04,GonzalezGarcia_05}.  In the
energy-independent oscillation mechanism that we are focusing on, the
extra contribution could be due to VCPT through Lorentz invariance
violation, in which case
\begin{equation}
\Delta \eta_{ij}^{\left(0\right)} = b_i-b_j \equiv b_{ij},
\end{equation}
with $b_i$ the eigenvalues of the Lorentz-violating CPT-odd
operator\cite{GonzalezGarcia_04} $\overline{\nu}_L
b_\mu^{\alpha\beta}\gamma_\mu\nu_L^\beta$. Alternatively, the
contribution could be due to NUCQ, and in this
case\cite{DeSabbata:1981ek} we would consider different couplings,
$k_i\neq k_j$ (for mass eigenstates $i$ and $j$), to a torsion field
$Q$, so that
\begin{equation}
\Delta \eta_{ij}^{\left(0\right)} 
= Q\left(k_i-k_j\right) \equiv Q k_{ij}.
\end{equation}
Strict bounds\cite{GonzalezGarcia_04} have been set on the parameters that control
the energy-independent NP mechanism using data from atmospheric and solar neutrinos, as well as SK and K2K, with energies up to about 1 TeV:
\begin{equation}
 b_{21} \le 1.6\times10^{-21} ~\text{GeV} ~~~, ~~~~ b_{32} \le 5.0\times10^{-23} ~\text{GeV} ~.
\end{equation}

Because the relative dominance of the NP energy-independent phase over
the standard oscillation phase increases with neutrino energy,
i.e. $\left(\Delta \tilde{p}_{ij} - \Delta p_{ij}\right) / \Delta
p_{ij} \sim E$, we would like to look at the most energetic neutrinos
available. Hence, we will consider neutrinos originating at cosmic
accelerators, such as active galactic nuclei, where it is presumed
that they are created with energies of up to $10^{11}$ GeV. Because
the typical distance to these accelerators is in the order of hundreds
of Mpc, we must include in the flavour-transition probability the
effect of cosmological expansion.  Hence, instead of the argument that
appears in the sine of Eq. (\ref{ProbKayserStdB}), we define an
accumulated phase\cite{Lunardini_01} $\phi_{ij}$ as follows:
\begin{eqnarray}\label{PhiStdWithT}
\phi_{ij}\left(t_f, t_i\right) = \int^{t_{f}}_{t_{i}} \Delta p_{ij}\left(\tau\right) d\tau = \int^{t_{f}}_{t_{i}} \frac{\Delta m_{ij}^2}{2E_{o}}\left(\frac{\tau}{t_{o}}\right)^{2/3} d\tau = \frac{3}{10} \frac{\Delta m_{ij}^2 t_{o}}{E_{o}} \left[\left(\frac{t_{f}}{t_{o}}\right)^{5/3} - \left(\frac{t_{i}}{t_{o}}\right)^{5/3}\right] ~,
\end{eqnarray}
where $t_{i}$ and $t_{f}$ are the times at which the neutrino was
produced and detected, respectively; $t_{o} = 13.7~\text{Gyr}$ is the
age of the Universe\cite{PDG_06}; and we have used the relation
between the energy at detection ($E_o$) and production epochs ($E$),
in an adiabatically expanding universe,
$E\left(\tau\right)=E_{o}\left(t_{o}/\tau\right)^{2/3}=E_{o}\left(1+z\right)$.
Considering the detection time $t_f$ in the present epoch, $t_{f} =
t_{o}$, we obtain the accumulated phase
\begin{equation}\label{PhiStd}
\phi_{ij}\left(E_{o}, z\right) = 1.97\times10^{23} \frac{\Delta m_{ij}^2\left[\text{eV}^2\right]}{E_{o}\left[\text{GeV}\right]} \left[1 - \left(1 + z\right)^{-5/2}\right] ~,
\end{equation}
where we have made use of the relation $t_{i}/t_{o} = \left(1 +
z\right)^{-3/2}$.

By replacing the momenta difference $\Delta p_{ij}$ with $\Delta
\tilde{p}_{ij}$, we obtain, correspondingly,
\begin{equation}\label{PhiNP}
\tilde{\phi}_{ij}\left(E_{o}, z\right) =  \phi_{ij}\left(E_o, z\right) + \frac{\Delta \eta_{ij}^{\left(n\right)}E_o^nt_o}{2} \left[1-\left(1+z\right)^{n-3/2}\right] \equiv \phi_{ij}\left(E_{o}, z\right) + \xi_{ij}^{\left(n\right)}\left(E_{o}, z\right) ~,
\end{equation}
with $\xi_{ij}^{\left(n\right)}$ the contribution to the phase due to
the NP effects. For $n=0$,
\begin{equation}\label{PhaseNPn0}
\xi_{ij}^{\left(0\right)}\left(E_o,z\right) = 3.28 \times 10^{41} 
b_{ij} \left[\text{GeV}\right] \left[1-\left(1+z\right)^{-3/2}\right] ~.
\end{equation}
Hence, instead of the traditional expression in Eq. (\ref{ProbKayserStdB}) for $P_{\nu_\alpha\rightarrow\nu_\beta}$, we will employ
\begin{equation}\label{ProbLun}
P_{\nu_\alpha\rightarrow\nu_\beta}\left(E_{o}, z\right) = \delta_{\alpha\beta} - 
4 \sum_i \text{Re}\left(J_{ij}^{\alpha\beta}\right) \sin^2\left(\frac{\phi_{ij}}{2}\right) ~,
\end{equation}
where the explicit expression for $\phi_{ij} \equiv
\phi_{ij}\left(E_{o}, z\right)$ is either of Eqs. (\ref{PhiStd}) or
(\ref{PhiNP}), depending on which dispersion relation is being
considered\footnote{In the limit of very small $z$,
Eq. (\ref{ProbLun}) reduces to Eq. (\ref{ProbKayserStdB}), the
expression for the transition probability for neutrinos that travel
much less than cosmological distances, i.e. solar, atmospheric and
reactor neutrinos. This can be seen by making $t_i=t_o-\Delta t$ in
Eq. (\ref{PhiStdWithT}), with $\Delta t \ll 1$, and discarding terms
of order $\left(\Delta t\right)^2$ and higher.}.

\section{Observability of the NP effects in the high-energy neutrino 
flavour ratios}\label{Section_Observability}

Using the flavour-transition probability obtained in the previous
section, Eq. (\ref{ProbLun}), we can calculate the ratio of neutrinos
of each flavour to the total number of neutrinos that arrive at the
detector from a source with redshift $z$. For $\alpha$-flavoured
neutrinos ($\alpha=e,\mu,\tau$) with energy $E_{o}$, this is
\begin{equation}\label{DetFlavourRatios}
\Upsilon_{\nu_\alpha}^{D}\left(E_{o},z\right) = 
\sum_{\beta = e, \mu, \tau}  P_{\nu_\beta\rightarrow\nu_\alpha}
\left(E_{o},z\right) \Upsilon_{\nu_\beta}^{S}  ~,
\end{equation}
where $\Upsilon_{\nu_\alpha}^{D}$ is the ratio at the detector and
$\Upsilon_{\nu_\beta}^{S}$ is the ratio at the source. The latter is
estimated assuming that neutrinos are secondaries of high-energy
proton-proton or proton-photon collisions, which produce pions that
decay into neutrinos and muons which decay into neutrinos
too\cite{Christian_05,Athar_00,Athar_06}:
$
\pi^{+} \longrightarrow \mu^{+}~\nu_{\mu} 
\longrightarrow e^{+}~\nu_{e}~\overline{\nu}_{\mu}~\nu_{\mu}~,~
\pi^{-} \longrightarrow \mu^{-}~\overline{\nu}_{\mu} 
\longrightarrow e^{-}~\overline{\nu}_{e}~\nu_{\mu}~\overline{\nu}_{\mu}  .
$
It is easy to see that
$\Upsilon_{\nu_e}^{S}:\Upsilon_{\nu_\mu}^{S}:\Upsilon_{\nu_\tau}^{S} =
1/3:2/3:0$. (Actually, $\nu_{\tau}$ \textit{are} expected to be
produced through the decay of $D_{s}^{\pm}$ charmed mesons generated
also in $pp$ and $p\gamma$ collisions. However, $D_{s}^{\pm}$
production is strongly suppressed\cite{Athar_06} and $\Upsilon_{\nu_\tau}^{S} <
10^{-5}$.)

The $\Upsilon_{\nu_\alpha}^D$ are very rapidly oscillating functions of
energy. Taking into account the limited
energy sensitivity of current and envisioned neutrino telescopes
(AMANDA-II, for instance, had an energy resolution of $0.4$ in the logarithm of
the energy of the $\nu_\mu$-spawned muon\cite{Ahrens:2004qq}), we see that
they are sensitive not to the instantaneous value of the ratios,
$\Upsilon_{\nu_\alpha}^D\left(E_o,z\right)$, but rather to the
energy-averaged flavour ratios
\begin{equation}\label{DetFlavourRatiosAvg}
\langle\Upsilon_{\nu_\alpha}^D\left(E_o,z\right)\rangle 
= \frac{1}{\Delta E_o}\int_{E_o^{min}}^{E_o^{max}} \Upsilon_{\nu_\alpha}^D\left(E_o',z\right)~dE_o' ~,
\end{equation}
where $E_o^{min}=E_o-\delta E_o$, $E_o^{max}=E_o+\delta E_o$ and $\Delta E_o\equiv E_o^{max}-E_o^{min} = 2 \delta E_o$, with $\delta E_o$ a small energy displacement.
Without the NP effects, the high-energy neutrino flux from a
distant astrophysical source is equally distributed among the three
flavours, i.e. $\langle\Upsilon_{\nu_e}^{D}\rangle:\langle\Upsilon_{\nu_\mu}^{D}\rangle:\langle\Upsilon_{\nu_\tau}^{D}\rangle
= 1/3:1/3:1/3$. 

\begin{figure}[t]
  \begin{center}
    \scalebox{0.5}{\includegraphics{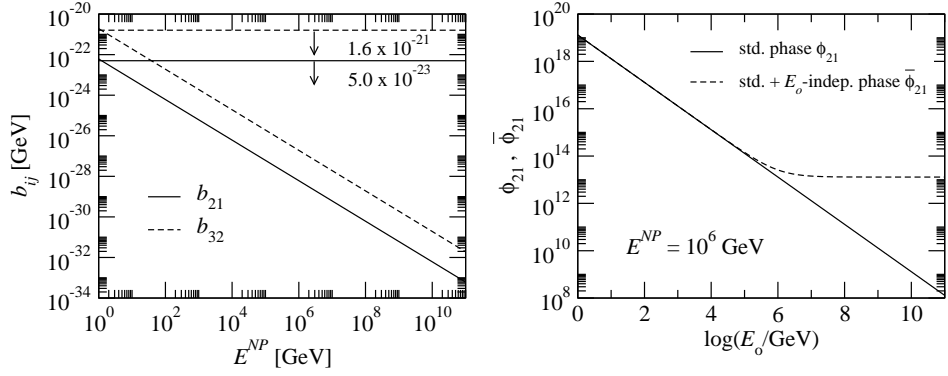}} 
     \caption{(left) Eigenvalues $b_{21}$ and $b_{32}$ as functions of
     $E^{NP}$, the energy at which the standard and NP
     energy-independent oscillation phases become comparable,
     i.e. $\phi_{ij} \sim \xi_{ij}^{\left(0\right)}$, according to
     Eq.~(\ref{Enpn0}). The redshift $z = 1$. The current upper bounds
     are plotted as horizontal lines. Notice that, due to these
     bounds, $E^{NP}$ cannot be lower than about 1 GeV.  (right)
     Standard oscillation phase $\phi_{21}$ and phase including the
     energy-independent contribution, $\tilde{\phi}_{21}$, as
     functions of neutrino energy. The redshift $z = 1$. Note that the
     phases start to differ at $E^{NP} = 10^6$ GeV, which corresponds
     to $b_{21} = 6.1 \times 10^{-29}$ GeV and $b_{32} = 1.9 \times
     10^{-27}$ GeV. Below this energy, they are indistinguishable.}
    \label{Fig1}
  \end{center}
\end{figure}

In the presence of NP effects, however, the detected flavour ratios
might be modified. Given that the relative dominance of the
energy-independent NP phase $\xi_{ij}^{\left(0\right)}$ over the
standard phase $\phi_{ij}$ grows with energy,
i.e. $\xi_{ij}^{\left(0\right)}/\phi_{ij}\sim E_o$, we would expect
that any modifications became more pronounced in the UHE range,
PeV--EeV, or higher. As explained in Section
\ref{Section_ModificationsQG}, while the NP phase remains constant in
energy, the standard phase decreases and, as a consequence, beyond a
certain threshold (determined by the values of the $b_{ij}$), the
detected ratios $\Upsilon_{\nu_\alpha}^D$ would acquire a constant
nonzero value, which might differ from the standard ratios
$1/3:1/3:1/3$, thus providing a distinct phenomenological signature of
a possible energy-independent contribution to the oscillation phase.

\begin{figure}[t]
  \begin{center}
    \scalebox{0.5}{\includegraphics{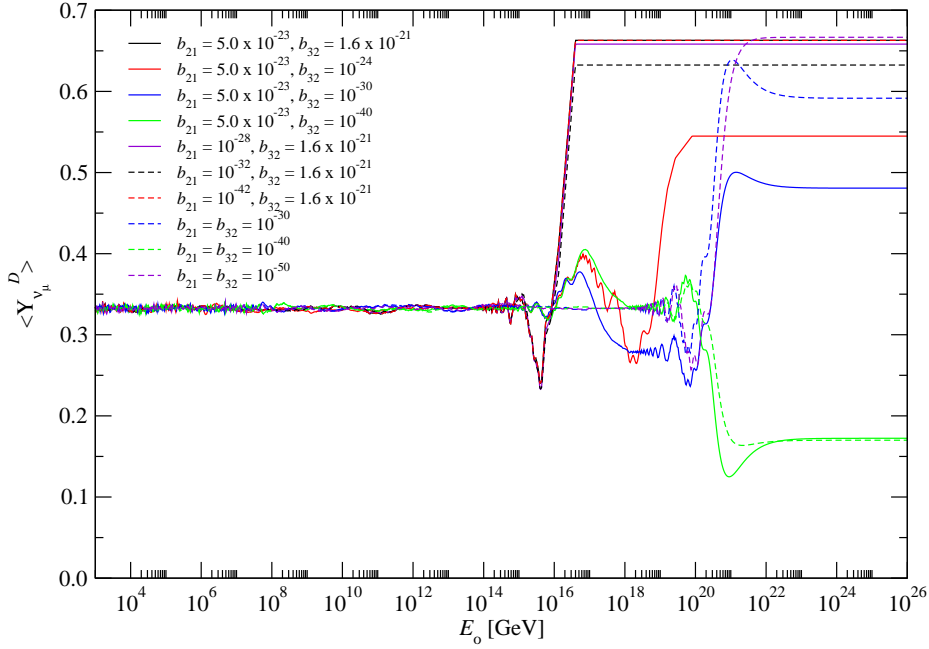}} 
     \caption{Energy-averaged detected $\nu_\mu$ ratio
     $\Upsilon_{\nu_\mu}^D$, Eq.~(\ref{DetFlavourRatiosAvg}), as a
     function of neutrino energy $E_o$, for different values of the
     $b_{ij}$. Note that the ratio becomes constant only for
     unrealistically high energies: $\sim 10^{16.5}$ GeV in the best
     case, when the $b_{ij}$ are set to their upper bounds. For lower
     values of the $b_{ij}$, the energy at which the ratio becomes
     constant is higher. The neutrino flux from cosmic accelerators is
     predicted to span up to about $10^{11}$ GeV; hence, the regime of
     constant $\Upsilon_{\nu_\mu}^D$ due to an energy-independent
     contribution to the oscillation phase would not be observable.}
    \label{FigNUMURATIOAVG}
  \end{center}
\end{figure}

As a means of estimating the values of the $b_{ij}$ for which the NP
phase starts to be of importance, we can demand that
$\xi_{ij}^{\left(0\right)} \sim \phi_{ij}$. From this requirement, we
can calculate, for given values of the $b_{ij}$, the energy $E^{NP}$
above which the NP effects are expected to become increasingly more
dominant in the oscillation. Doing this, we obtain
\begin{equation}\label{Enpn0}
E^{NP} \left[\text{GeV}\right] = 6 \times 10^{-19} 
\frac{\Delta m_{ij}^2 \left[\text{eV}^2\right]}
{b_{ij} \left[\text{GeV}\right]}
\frac{1-\left(1+z\right)^{-5/2}}{1-\left(1+z\right)^{-3/2}} ~.
\end{equation}
The left panel of Fig.~\ref{Fig1} shows a plot of $b_{21}$ and $b_{32}$
as functions of $E^{NP}$. The current upper bounds are shown as
horizontal lines. The lower the value of $E^{NP}$, the earlier the NP
effects would manifest. Notice that, due to the current bounds,
$E^{NP}$ cannot be lower than about 1 GeV. The plots have been
generated for a fixed $z=1$; for lower values of $z$ we will have a higher 
value of $b_{ij}$ ($30\%$ if we take $z = 0.03$), while we will obtain a 
decrease in the values of $b_{ij}$ for large $z$ (a $20\%$ 
decrease for $z = 6$).
The right panel of Fig.~\ref{Fig1} shows the
standard and the energy-independent NP phases, $\phi_{21}$ and
$\tilde{\phi}_{21}$, respectively, as functions of neutrino energy,
assuming that $E^{NP} = 10^6$ GeV. Notice that the phases start to
differ precisely at this energy.

For concreteness, we will study the detected ratio of $\nu_\mu$
defined in Eq.~(\ref{DetFlavourRatiosAvg}), since our conclusions are
independent of the chosen flavour. Fig.~\ref{FigNUMURATIOAVG} shows
the predicted ratio calculated for different values of $b_{21}$ and
$b_{32}$ (we have assumed that $b_{31} = b_{32} + b_{21}$) as a
function of $E_o$.

Fig.~\ref{FigNUMURATIOAVG} shows that $\Upsilon_{\nu_\mu}^D$ indeed
becomes constant and different from $1/3$ after a certain energy
threshold. This occurs when the standard phase $\phi_{ij} \rightarrow
0$, so that, effectively, the oscillation phase is reduced to the
energy-independent NP contribution, i.e. $\tilde{\phi}_{ij}
\rightarrow \xi_{ij}^{\left(0\right)}$, and the transition
probabilities become constant. Note, however, that
$\Upsilon_{\nu_\mu}^D$ is constant only for $E_o \gtrsim 10^{16.5}$
GeV in the most promising case, that is, when the $b_{ij}$ equal their
current upper bounds. This is about five orders of magnitude higher
than the energy of the most energetic neutrinos expected from cosmic
accelerators. For smaller values of these parameters, the energy at
which the ratio becomes constant is even higher. Using closer or more
distant sources, effectively decreasing or increasing $z$, does not
affect the energy threshold, but only modifies the constant value
reached by $\Upsilon_{\nu_\mu}^D$. Therefore, we conclude that, given
the current upper bounds on the $b_{ij}$, an energy-independent NP
contribution to neutrino oscillations would be visible in the
high-energy astrophysical neutrino flux only if it modifies the
oscillation amplitude (i.e. the mixing angles), as well as the phase.

In light of this conclusion, within the formalism used in the present
work, a comparative calculation, with and without NP effects, of
high-energy astrophysical neutrinos detected at a second-generation
neutrino telescope such as IceCube becomes unnecessary.

\section{Summary and conclusions}\label{Section_Conclusions}

We have considered the effect of a modified dispersion relation on the
detected flavour ratios of high-energy neutrinos from cosmic
accelerators. In the scenario of new physics that we have explored,
the flavour oscillation phases are modified by the addition of
energy-independent terms which depend on the parameters $b_{ij}$. This
contribution could correspond to a violation of CPT symmetry or to a
nonuniversal neutrino coupling to a torsion field. The current upper
bounds on the $b_{ij}$ are strict: $b_{21} \le 5.0 \times 10^{-23}$
GeV and $b_{32} \le 1.6 \times 10^{-21}$ GeV.

At sufficiently high energies, the oscillation phases are dominated by
the energy-independent terms and the flavour ratios become constant
and, possibly (depending on the values of the $b_{ij}$) different
from the average value of the ratios in the standard oscillation case,
when new physics effects are absent. We have found, however, that even
in the best case, when the $b_{ij}$ are set to their upper bounds, the
ratios are constant only for energies above $10^{16.5}$ GeV, about
five orders of magnitude higher than the most energetic neutrinos that
are expected from cosmic accelerators. Lower values of the $b_{ij}$
will only result in higher energy thresholds for the ratios to become
constant.

Therefore, we conclude that, even though there could be, in principle,
a clear signature of the presence of energy-independent contributions
to the neutrino flavour oscillations, these are not detectable in the
flavour ratios of high-energy neutrinos from cosmic accelerators if
they affect solely the oscillation phases.

\section*{Acknowledgments}

This work was supported by grant number 3256 from
the Direcci\'on Acad\'emica de Investigaci\'on of the Pontificia
Universidad Cat\'olica del Per\'u, by CONACyT M\'exico, and by a 
High Energy Physics Latin American European Network (HELEN) Type AT 
Grant. MB would like to thank the Physics
Department at Centro de Investigaci\'on y de Estudios Avanzados del IPN
(CINVESTAV) for its hospitality during the development of this work.

\end{document}